\documentclass[%
 reprint,
superscriptaddress,
bibnotes,
 amsmath,amssymb,
 aip,
]{revtex4-2}

\usepackage{graphicx}
\usepackage{dcolumn}
\usepackage{bm}
\usepackage{hyperref}
\usepackage{natmove}
\usepackage{color}

\begin{document}

\preprint{APS/123-QED}

\title{Dissipation bounds the amplification of transition rates far from equilibrium}

\author{Benjamin Kuznets-Speck}
\affiliation{Biophysics Graduate Group, University of California, Berkeley, CA, 94720, USA \looseness=-1}
\author{David T. Limmer\email{dlimmer@berkeley.edu}}%
\affiliation{Chemistry Department, University of California, Berkeley, CA, 94720, USA \looseness=-1}
\affiliation{Chemical Sciences Division, Lawrence Berkeley National Laboratory, Berkeley, CA, 94720, USA \looseness=-1}
\affiliation{Material Sciences Division, Lawrence Berkeley National Laboratory, Berkeley, CA, 94720, USA \looseness=-1}
\affiliation{Kavli Energy NanoSciences Institute, University of California, Berkeley, CA, 94720, USA \looseness=-1}

\date{\today}


\maketitle


\renewcommand{\vec}[1]{\mathbf{#1}}
\newcommand{\n}{\mathrm{neq}}
\newcommand{\e}{\mathrm{eq}}
\newcommand{\kB}{k_\mathrm{B}}

\newcommand{\dU}{\Delta U}
\newcommand{\dUt}{\Delta \tilde U}
\newcommand{\dUh}{\Delta \hat U}
\newcommand{\dUn}{\left \langle  \Delta U \right \rangle^\mathrm{neq}}
\newcommand{\dUe}{\left \langle  \Delta U \right \rangle^\mathrm{eq}}
\newcommand{\Un}{U^\mathrm{neq}}
\newcommand{\Ue}{U^\mathrm{eq}}

\newcommand{\MGneqAB}{\left \langle e^{\Delta U} \right \rangle^\mathrm{neq}_{AB}}
\newcommand{\MGneqAA}{\left \langle e^{\Delta U} \right \rangle^\mathrm{neq}_\mathrm{AA}}
\newcommand{\MGeqAB}{\left \langle e^{-\Delta U} \right \rangle^\mathrm{eq}_{AB}}
\newcommand{\MGneqA}{\left \langle e^{\Delta U} \right \rangle^\mathrm{neq}_A}
\newcommand{\MGeqAA}{\left \langle e^{-\Delta U} \right \rangle^\mathrm{eq}_\mathrm{AA}}
\newcommand{\MGeqA}{\left \langle e^{-\Delta U} \right \rangle^\mathrm{eq}_A}
\newcommand{\MGeqi}{\left \langle e^{-\Delta U} \right \rangle^\mathrm{eq}_i}
\newcommand{\MGneqi}{\left \langle e^{\Delta U} \right \rangle^\mathrm{neq}_i}

\newcommand{\MGneqABIC}{\left \langle e^{\Delta U+\ln \frac{\rho_\mathrm{eq}}{\rho_\mathrm{neq}}} \right \rangle^\mathrm{neq}_{AB}}
\newcommand{\MGneqAIC}{\left \langle e^{\Delta U+\ln \frac{\rho_\mathrm{eq}}{\rho_\mathrm{neq}}} \right \rangle^\mathrm{neq}_A}
\newcommand{\MGeqABIC}{\left \langle e^{-\Delta U-\ln \frac{\rho_\mathrm{eq}}{\rho_\mathrm{neq}}} \right \rangle^\mathrm{eq}_{AB}}
\newcommand{\MGeqAIC}{\left \langle e^{-\Delta U-\ln \frac{\rho_\mathrm{eq}}{\rho_\mathrm{neq}}} \right \rangle^\mathrm{eq}_A}

\newcommand{\dUnAB}{\left \langle \Delta U \right \rangle^\mathrm{neq}_{AB}}
\newcommand{\dUnAA}{\left \langle \Delta U \right \rangle^\mathrm{neq}_\mathrm{AA}}
\newcommand{\dUnA}{\left \langle \Delta U \right \rangle^\mathrm{neq}_A}
\newcommand{\dUeAB}{\left \langle \Delta U \right \rangle^\mathrm{eq}_{AB}}
\newcommand{\dUeAA}{\left \langle \Delta U \right \rangle^\mathrm{eq}_\mathrm{AA}}
\newcommand{\dUeA}{\left \langle \Delta U \right \rangle^\mathrm{eq}_A}

\newcommand{\ddUnAB}{\left \langle \delta \Delta U \right \rangle^\mathrm{neq}_{AB}}
\newcommand{\ddUeAB}{\left \langle \delta \Delta U \right \rangle^\mathrm{eq}_{AB}}

\newcommand{\Oe}{\left \langle O \right \rangle^\mathrm{eq}}
\newcommand{\On}{\left \langle O \right \rangle^\mathrm{neq}}

\newcommand{\MGneq}{\left \langle e^{\Delta U} \right \rangle^\mathrm{neq}}
\newcommand{\MGeq}{\left \langle e^{-\Delta U} \right \rangle^\mathrm{eq}}
\newcommand{\MGOneq}{\left \langle O e^{\Delta U} \right \rangle^\mathrm{neq}}
\newcommand{\MGOeq}{\left \langle O e^{-\Delta U} \right \rangle^\mathrm{eq}}

\newcommand{\dUdotnAB}{\left \langle \Delta \dot U \right \rangle^\mathrm{neq}_{AB}}
\newcommand{\dUdotnA}{\left \langle \Delta \dot U \right \rangle^\mathrm{neq}_A}
\newcommand{\dUdoteAB}{\left \langle \Delta \dot U \right \rangle^\mathrm{eq}_{AB}}
\newcommand{\dUdoteA}{\left \langle \Delta \dot U \right \rangle^\mathrm{eq}_A}

\newcommand{\pn}{P^\mathrm{neq}[\vec X(t)]}
\newcommand{\pe}{P^\mathrm{eq}[\vec X(t)]}
\newcommand{\phn}{\hat P^\mathrm{neq}}
\newcommand{\phe}{\hat P^\mathrm{eq}}
\newcommand{\ptn}{\tilde P^\mathrm{neq}}
\newcommand{\pte}{\tilde P^\mathrm{eq}}
\newcommand{\rhon}{\rho_\mathrm{neq}}
\newcommand{\rhoe}{\rho_\mathrm{eq}}

\newcommand{\QeA}{\beta \left \langle Q \right \rangle^\mathrm{eq}_A}
\newcommand{\QnA}{\beta\left \langle Q \right \rangle^\mathrm{neq}_A}
\newcommand{\QeAB}{\beta\left \langle Q \right \rangle^\mathrm{eq}_{AB}}
\newcommand{\QnAB}{\beta\left \langle Q \right \rangle^\mathrm{neq}_{AB}}

\newcommand{\AeA}{\left \langle A \right \rangle^\mathrm{eq}_A}
\newcommand{\AnA}{\left \langle A \right \rangle^\mathrm{neq}_A}
\newcommand{\AeAB}{\left \langle A \right \rangle^\mathrm{eq}_{AB}}
\newcommand{\AnAB}{\left \langle A \right \rangle^\mathrm{neq}_{AB}}

\newcommand{\kamp}{\frac{k_\lambda}{k_0}}
\newcommand{\kampo}{k_\lambda/k_0}

\newcommand{\iAAB}{\mathrm{i} \in \{A,\mathrm{B} \}}

\newcommand{\ZA}{Z_A}
\newcommand{\ZAB}{Z_AB}
\newcommand{\ZnA}{Z^\mathrm{neq}_A}
\newcommand{\ZnAB}{Z^\mathrm{neq}_{AB}}
\newcommand{\ZeA}{Z^\mathrm{eq}_A}
\newcommand{\ZeAB}{Z^\mathrm{eq}_{AB}}
\newcommand{\Zni}{Z^\mathrm{neq}_i}
\newcommand{\Zei}{Z^\mathrm{eq}_i}

\newcommand{\cAA}{C_\mathrm{AA}(t)}
\newcommand{\cAAn}{C_\mathrm{AA}^\mathrm{neq}(t)}
\newcommand{\cAAe}{C_\mathrm{AA}^\mathrm{eq}(t)}
\newcommand{\cAB}{C_AB(t)}
\newcommand{\ps}{p_s^\mathrm{neq}(t)}
\newcommand{\psn}{p_s^\mathrm{neq}(t)}
\newcommand{\pse}{p_s^\mathrm{eq}(t)}

\newcommand{\keq}{k_\mathrm{eq}}
\newcommand{\kneq}{k_\mathrm{neq}}
\newcommand{\kAB}{k_{AB}}
\newcommand{\lam}{\lambda}

{\bf Complex systems can convert energy imparted by nonequilibrium forces to regulate how quickly they transition between long lived states. While such behavior is ubiquitous in natural and synthetic systems, currently there is no general framework to relate the enhancement of a transition rate to the energy dissipated, or to bound the enhancement achievable for a given energy expenditure. 
We employ recent advances in statistical thermodynamics to build such a framework, which can be used to gain mechanistic insight on transitions far from equilibrium. We show that under general conditions, there is a basic speed-limit relating the typical excess heat dissipated throughout a transition and the rate amplification achievable. We illustrate this trade-off in canonical examples of diffusive barrier crossings in systems driven with autonomous and deterministic external forcing protocols. In both cases, we find that our speed limit tightly constrains the rate enhancement. }                 

The natural world is full of systems in which the rate of a rare dynamical event is enhanced through coupling to a dissipative process.~\cite{carlos_review, mike_mm_review} 
\emph{In vivo}, molecular chaperones accelerate protein folding and assembly so that otherwise slow transitions occur on biologically relevant timescales, at the energetic cost of maintaining chemical potential gradients~\cite{chaperone_review_2}. Shear forces drive colloidal assemblies and polymer films to order rapidly enough for viable synthesis, at the expense of applying external forces~\cite{colloid_shear}. Such behavior is leveraged across physical and biological systems, but there are few known principles available to act as guides or constrain possibilities. Here we use nonequilibrium stochastic thermodynamics to demonstrate that dissipation bounds the enhancement of the rate of a transition away from equilibrium. The bound is sharp near equilibrium and for large barriers, holds arbitrarily far from equilibrium, and can be tightened with additional knowledge of kinetic factors. Our work thus elucidates a fundamental trade-off between speed and energy consumption. 

In equilibrium, the rate of a transition between two long lived states is determined by the likelihood that a thermal fluctuation provides sufficient energy to the system to overcome a free energy barrier. Away from equilibrium, external forces and nonthermal fluctuations can mitigate this constraint, modulating the rate relative to its equilibrium value.  Departures from thermal equilibrium make it difficult to predict the extent to which a dissipative process can influence a transition, as traditional rate theories are grounded in equilibrium statistical mechanics. For instance, both classical transition state theory~\cite{TST} and Kramer's theory ~\cite{kramers} require information on the probability to reach a rare dividing surface, or transition state. In equilibrium the Boltzmann distribution supplies that probability, but within a nonequilibrium steady-state that information is generally unavailable. Freidlin-Wentzell theory~\cite{Boucher}, and transition path theory~\cite{TPS_review} supply formal means of estimating rates away from equilibrium through the consideration of path ensembles. However, rate calculations within these formalisms require complex optimizations or partition function evaluations, and do not encode simple relationships between rates and other measurable quantities.

Using principles from stochastic thermodynamics, we develop a general theory of nonequilibrium rate enhancement, deriving exact relations and fundamental bounds\cite{seifert_review}. Stochastic thermodynamics has supplied a number of relationships that constrain fluctuations away from equilibrium in terms of measurable energetic observables~\cite{Jarzynski_equality}. The fluctuation theorems illustrate fundamental time-reversal symmetries~\cite{CrooksFT}, and thermodynamic uncertainty relations bound
response~\cite{TUR_Todd_review}.  In this work, we show that the rate enhancement achievable away from equilibrium is bounded by the heat dissipated over the course of the transition, 
\begin{equation}
\label{Eq:main}
\frac{\kneq}{\keq} \leq  e^{\beta \bar{\mathcal{Q}} /2}  ,
\end{equation}
where $\kneq/\keq$ is the ratio of the nonequilibrium to equilibrium transition rates, and deviation from equilibrium due to broken detailed balance is codified by $\bar{\mathcal{Q}}$, the average excess heat released over the transition due to the nonequilibrium process, in units of $\kB T$, where $\kB$ is Boltzmann's constant and $T$ the temperature of the bath.  Our theory demonstrates that the rate enhancement achievable by coupling a system to a dissipative process, an essential dynamical quantity, is limited by general thermodynamic constraints. To test the theory, we study paradigmatic two-state continuous force systems, driving them from equilibrium with both deterministic and autonomous forces.  

\section*{Stochastic thermodynamics of rate enhancement}
To derive Eq. \ref{Eq:main}, we consider systems driven by a time-dependent force, $\vec{\lambda}(t)$, either externally controlled or coupled to an additional nonthermal noise-source that evolves independently of the system-state, precluding feedback. Extensions to systems evolving in boundary driven nonequilibrium steady-states, though likely possible, are not explored in this work.

In the presence of the time-dependent force, the rate, $k_{\lambda}$, of transition between two long-lived states, is the probability that a transition occurs per unit time. For a system described by a configuration, $\vec x(t)$, at time $t$, we will consider initial and final states, $A$ and $B$, that are collections of configurations defined by the indicator functions,
\begin{equation}
  h_i(t) =
    \begin{cases}
      1 & \text{if $\vec x(t) \in i$}\\
      0 & \text{else}
    \end{cases}       
\end{equation}
where $i \in \{ A,B \}$, and we assume $A$ and $B$ are not intersecting. 
For times longer than the characteristic local relaxation time and much shorter than the inverse rate, $k_\lambda$ derives from a ratio of path partition functions,
\begin{equation}
\label{Eq:rate}
 k_{\lambda}(A\rightarrow B)  =   \frac{d}{d t} \frac{Z_{AB}(\lambda)}{Z_A(\lambda)}. 
\end{equation}
Here,
\begin{equation}
 Z_{AB}(\lambda) = \int D[\vec X(t)] h_A(0) h_\mathrm{B}(t) P_{\lambda}[\vec X(t)] 
\end{equation}
is the number of transition paths, $\vec X (t) =\{ \vec x(0),...,\vec x(t) \}$, starting in $A$ and ending in $B$ at time $t$, weighted with probability $P_\lambda[\vec X(t)]$, and
\begin{equation} 
 Z_{A}(\lambda) =\int D[\vec X(t)] h_A(0) P_{\lambda}[\vec X(t)]
\end{equation}
is the corresponding number of paths starting in $A$ \cite{TPS_review}. 

The ratio in Eq.~\ref{Eq:rate} is simply the conditional probability of the system being in state $B$ given it started in $A$.  Provided the transition is rare, consistent with $A$ and $B$ representing metastable states, there is a range of time over which $Z_{AB}(\lambda)$ increases linearly, and $k_{\lambda}$ is constant. Specifically, the rate constant is defined for observation times $\tau_A \lesssim t \ll k_\lambda^{-1}$, where the transition path time is typically on the order of $\tau_\mathrm{A}$,  the characteristic relaxation time within state $A$, and shorter than the timescale required for global relaxation.  The probability of a path is the product of a distribution of initial conditions, $\rho_\lambda[\vec x(0)]$, and the conditional transition probability $P_\lambda[\vec X(t)| \vec x(0)]$, such that $P_\lambda[\vec X(t)] = P_\lambda[\vec X(t)| \vec x(0)] \rho_\lambda[\vec x(0)]$. While in general away from thermal equilibrium, $\rho_\lambda[\vec x(0)]$ is unknown, $P_\lambda[\vec X(t)| \vec x(0)]$ can be inferred, provided an equation of motion. For the specific model calculations discussed below, $P_\lambda[\vec X(t)| \vec x(0)]$ will take an Onsager-Machlup form\cite{Onsager_Machlup}.
 
Stochastic thermodynamics gives structure to path ensembles and relations to thermodynamic quantities. In an equilibrium system, the principle of microscopic reversibility implies that the probability of a trajectory is equal to its time-reverse. Specifically, let $\tilde{P}_{\lambda}[\tilde{\vec X}(t)]$ denote the probability of observing a time-reversed trajectory $\tilde{\vec X}(t) =\{ \tilde{\vec x}(t),...,\tilde{\vec x}(0) \}$, where $\tilde{\vec x}(t)$ is a time-reversed configuration of the system at $t$, labeled in the forward time direction. In the absence of the dissipative protocol, $\lambda=0$, the system is in equilibrium and ${P}_0[{\vec X}(t)]=\tilde{P}_0[\tilde{\vec X}(t)]$. The Crooks fluctuation theorem extends this notion to systems driven away from equilibrium by an arbitrary time dependent force $\lambda(t)$\cite{CrooksFT}. For a nonequilibrium system, microscopic reversibility is manifested by 
${P}_{\lambda}[{\vec X}(t)| \vec x(0)]=\tilde{P}_{\lambda}[\tilde{\vec X}(t)| \vec x(t)] \exp[\beta (Q [{\vec X}(t)|\vec x(0)]+Q_\mathrm{rev} [{(\vec x(t),\vec x(0)]) ]}$.
The first term in the exponential, $\beta Q$, is what we refer to as \emph{dissipation}, as it is the excess heat transferred from the system to the bath along a trajectory driven from equilibrium over the corresponding heat transferred for a reversible process. For an equilibrium system, the reversible contribution to the heat is equal to the Shanon entropy, $\beta Q_\mathrm{rev} = \ln \rho_0[\vec x(t)]/\rho_0[\vec x(0)]$. It is generally derivable as the change in energy due to the conservative forces, and thus depends on only the trajectory's boundaries. 
  
Coupling the system to a dissipative process will generally change its dynamics. Using trajectory reweighting, we relate the transition rate in the presence and absence of the nonequilibrium force $\vec \lambda(t)$. We consider two path probability distributions with support on the same $\vec X(t)$, so that the relative action    
\begin{equation}
    \beta \dU_\lambda [{\vec X}(t)|{\vec x}(0)] =  \ln \frac{{P}_{\lambda}[{\vec X}(t)|{\vec x}(0)]}{{P}_{0}[{\vec X}(t)|{\vec x}(0)]} \, ,
\end{equation}
relating one to the other, is well-defined. Performing a change of measure, for a constant distribution of initial conditions, we express ratios of path partition functions in either ensemble as
\begin{equation}
\label{Eq:Reweight}
\frac{ Z_{AB}(\lambda)}{ Z_{AB}(0)} =\left \langle e^{\beta \dU_\lambda} \right \rangle_0 =\left \langle e^{-\beta \dU_\lambda} \right \rangle_\lambda^{-1} \, ,
\end{equation}
where the brackets denote a conditional average in a transition path ensemble connecting states $A$ and $B$ in time $t$, with path probability $P_0[\vec X(t)]$ in the first equality, or $P_\lambda[\vec X(t)]$ in the second equality. 

When transitions in both path ensembles are rare, $k_\lambda t$ and $k_0 t \ll 1$, the overwhelming majority of paths originating from $A$ will remain there on the timescales where the rate is time independent, so that $Z_A(\lambda)=Z_A(0)$. In Sec. S1., we consider generalizations away from this limit, showing that the ratio $Z_A(0)/Z_A(\lambda)$ cancels contributions in Eq.~\ref{Eq:Reweight} due to different distributions of initial conditions. Thus, under mild assumptions, combining Eq.~\ref{Eq:rate} with Eq.~\ref{Eq:Reweight}, we find 
\begin{equation}
    \frac{k_\lambda}{k_0} = \left \langle e^{\beta \dU_\lambda} \right \rangle_0 =\left \langle e^{-\beta \dU_\lambda} \right \rangle_\lambda^{-1}
\end{equation}
which is an exact relation between transition rates in the presence or absence of the dissipative process. Lower and upper bounds can be read off by applying Jensen's inequality to each of these expressions,
\begin{equation}
\label{Eq:Con}
\beta \left \langle  \dU_\lambda \right \rangle_0 \le \ln \frac{k_\lambda}{k_0}  \le \beta \left \langle  \dU_\lambda \right \rangle_\lambda 
\end{equation}
constituting a fundamental envelope for the rate enhancement. This result is general provided any two trajectory ensembles have common support. In a suitably defined linear response regime, the ensembles are approximately equal, $\left \langle  \dU_\lambda \right \rangle_\lambda \approx \left \langle  \dU_\lambda \right \rangle_0$, so the bounds are saturated. This corresponds to a near-equilibrium regime where driving is small.  While Eq.~\ref{Eq:Con} is derived for constant initial conditions for simplicity of notation, the impact of different initial conditions on an upper bound is to add a positive constant equal to a symmetrized Kullback–Leibler divergence between initial distributions in the driven and equilibrium ensembles (Sec. S1.). In the linear response regime, and when transition paths are long enough to lose memory, a common occurrence for rare transitions which quickly relax in both $A$ and $B$, changes to initial conditions can be neglected. 

Generally, $\Delta U_\lambda$ contains thermodynamic and kinetic factors. To separate them, we decompose $\Delta U_\lambda$ into time-reversal-symmetric and asymmetric trajectory observables,
\begin{equation}
  \Delta U_\lambda  = \frac{ Q+\Gamma}{2} ,
\end{equation}
where the excess heat $ Q = \dU_\lambda [{\vec X}(t)|{\vec x}(0)] - \dU_\lambda [\tilde{\vec X}(t)|\tilde{\vec x}(0)]$ is odd under time-reversal, and the excess dynamical activity $\Gamma=\dU_\lambda [{\vec X}(t)|{\vec x}(0)] + \dU_\lambda [\tilde{\vec X}(t)|\tilde{\vec x}(0)]$ is even.  On the whole, both the heat and the activity play important roles in response and stability of nonequilibrium systems~\cite{Maes_Frenesy, Chloe_David}.
While the heat has a simple mechanical definition and is largely independent of the system's dynamics, the activity depends on details of the equation of motion, making it hard to generalize. 

We find that for rare transitions across a host of physically relevant conditions, the activity can be neglected. Near equilibrium, the average activity in the conditioned transition ensemble vanishes due to time-reversal symmetry. In cases of instantonic transitions, where the driving force varies slowly relative to the characteristic transition path time the activity is small. 
Even away from limiting cases where the barrier is much larger than the scale of the noise, the activity can be neglected. 
Consider for instance free diffusion, where rare transitions are largely noise assisted. In these sojourns over broad, diffusive regions, the activity is strictly negative. By neglecting it, a bound based on the heat alone is satisfied though weakened. Each case is considered explicitly in Sec. S2. 
Remarkably, this implies that the dissipation accumulated over a transition bounds the rate enhancement,
\begin{equation} \label{diss_bound}
\ln \frac{k_\lambda}{k_0}  \le \frac{\beta}{2} \left \langle  Q \right \rangle_\lambda,
\end{equation}
which is our main result. Identifying the system under finite $\lambda$ as a nonequilibrium system, and its absence as an equilibrium one, we identify Eq.~\ref{diss_bound} as a more precise statement of Eq.~\ref{Eq:main}. We note, however that the rate enhancement relation is general for any two transition path ensembles.

\begin{figure}
\includegraphics[width=8.5 cm]{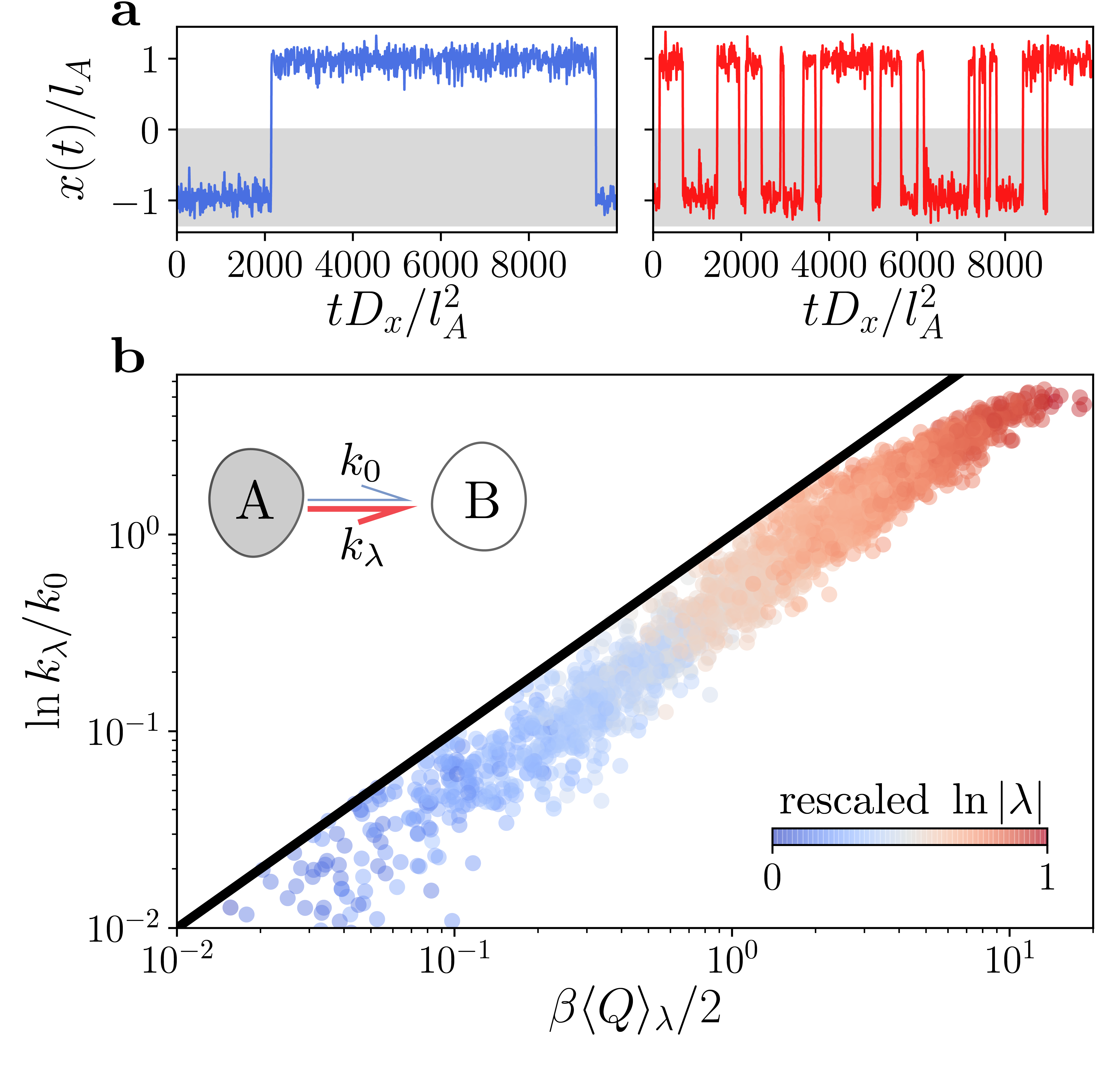}
\caption{\label{fig:1} Nonequilibrium driving enhances transition rates. $\mathbf{a}$, Trajectories of a two-state system in equilibrium (left) and the same process driven away from equilibrium (right) by a time-dependent external force. $\mathbf{b}$, Rate enhancement for different asymmetric two-state systems as a function of dissipated heat, each driven by a randomly chosen combination of deterministic and stochastic external forces. The bound in Eq. \ref{diss_bound} is shown as a black line and points are colored according to the magnitude of the force, low (blue) to high (red).}
\end{figure}

\section*{Autonomous and deterministic forcing}
To illustrate the robustness of our dissipative bound, we first consider the overdamped dynamics of a particle in a one-dimensional asymmetric potential subject to both external time dependent and nonthermal forces. Specifically, the equation of motion for the position of the particle, $x$, is taken as
$
\gamma \dot x = -\partial_x V(x) + \lambda(t) + \sqrt{2 \kB T \gamma} \eta_x
$
where $\gamma$ is the friction due to the surrounding medium, imposing a diffusion constant $D_x=\kB T /\gamma$, and $\eta_x$ is a Gaussian random variable with $\langle \eta_x(t)\rangle=0$ and $\langle \eta_x(t) \eta_x(t') \rangle= \delta (t-t')$. The static external potential consists of two quartic states, $V(x) = V_A(x) \Theta(-x)+V_B(x) \Theta(x)$. Each basin $V_{i}(x) =  (\Delta V_{i} x^2/(2 l_{i}^2)) (x^2/(2 l_{i}^2)-1)$, $i \in \{A,B\}$, is characterized by the distance of its minimum $(|x|,|y|) = (l_{i}, \Delta V_{i} )$ to the origin, where the states are joined by $\Theta(x)$, the Heaviside function. $V(x)$ supports two metastable states with a barrier between them if $\beta \Delta V_{i}>1$ for both $i=A$ and $B$. Figure.~\ref{fig:1}a), where $x(t)$ exhibits fluctuations concentrated around two regions of the potential, with few, fleeting transitions between them, manifests this metastability.  

We drive the system out of equilibrium according to a time-dependent protocol $\lambda(t) = f [ p \cos (\omega t) + (1-p) \cos (\theta(t))]$ with maximum amplitude $f$ partitioned, $p \in [0,1]$, into deterministic and autonomous components. The deterministic portion of the driving is periodic with frequency $\omega$, whereas the autonomous piece is determined by an additional nonthermal process, $\dot{\theta}(t) = \sqrt{2 D_\theta} \eta_\theta$, with diffusion constant $D_\theta$, and delta-correlated white noise, $\langle \eta_\theta(t)\rangle=0$ and $\langle \eta_\theta(t) \eta_\theta(t') \rangle= \delta (t-t')$. 

Considering transitions that take the particle from one side of the potential to the other, we define $h_A = \Theta(x+l_{A}/\sqrt{3})$ and $h_B=\Theta(x-l_{B}/\sqrt{3})$, which correspond to the locations of the maximum force opposing the transition in the absence of $\lambda(t)$. The dissipated excess heat can be computed from
\begin{equation}
    Q(t) =   \int^{t}_0 dt' \lambda(t) \dot x(t'), \label{Q_heat}
\end{equation}
and its mean estimated within the nonequilibrium steady-state by integrating the dissipation rate over reactive trajectories of length $t$, given by the typical transition path time as discussed in \emph{Methods} and Sec. S3. Given a suitable separation between inverse rate and relaxation time, $Q$ is often insensitive to the precise value of $t$. We reiterate that $Q$ differs from the total heat by the conservative boundary portion which appears in both the equilibrium and driven actions, and therefore does not play a role in the rate enchancement. 

Figure~\ref{fig:1}b) shows the results of 3000 randomly constructed models, where $\Delta V_{i}, l_{i}$, $p$, $f$, and $D_\theta$, were chosen uniformly over a wide range of parameters, as detailed in \emph{Methods}. For each model, $k_\lambda$, $k_0$ and $\langle Q \rangle_\lambda$ have been independently evaluated, and Fig.~\ref{fig:1}b) demonstrates that the bound holds across the broad parameter-space. Points are colored blue to red in increasing magnitude of the driving force $f$, showing that the protocol most efficiently amplifies the equilibrium rate when $\beta \langle Q \rangle_\lambda < 10$. Pushing past this regime, the bound becomes progressively weaker, as dissipation increases, but rate enhancement reaches a plateau. This corresponds to a limit where driving is large enough to degrade the assumption that basin $A$ is metastable.

In order to understand the physical processes that determine whether or not the bound is saturated, we focus on two cases of the model presented above. First, we set $p=0$, which corresponds to an active Brownian particle in an external potential. Active Brownian particles provide a canonical realization of how autonomous athermal noise can drive novel steady-states without simply imparting an effective temperature\cite{ABP_effective_temp}. These self-propelled agents exhibit dynamical symmetry breaking and collective motion \cite{optimizing_active_work,Trevor}, and previous studies have shown that the escape of active particles from a metastable potential exhibits interesting behavior arising from an interplay between the driving force, persistence time statistics, and the shape of the potential\cite{active_escape, woillez2020nonlocal}. For simplicity we take a symmetric potential, with $l_{A}=l_{B}=1$ and $\beta \Delta V_A=\beta \Delta V_B=10$, setting $\gamma=1$ and $\kB T=  1/2$. 

\begin{figure}[t]
\includegraphics[width=.925\linewidth, height=1.36\linewidth]{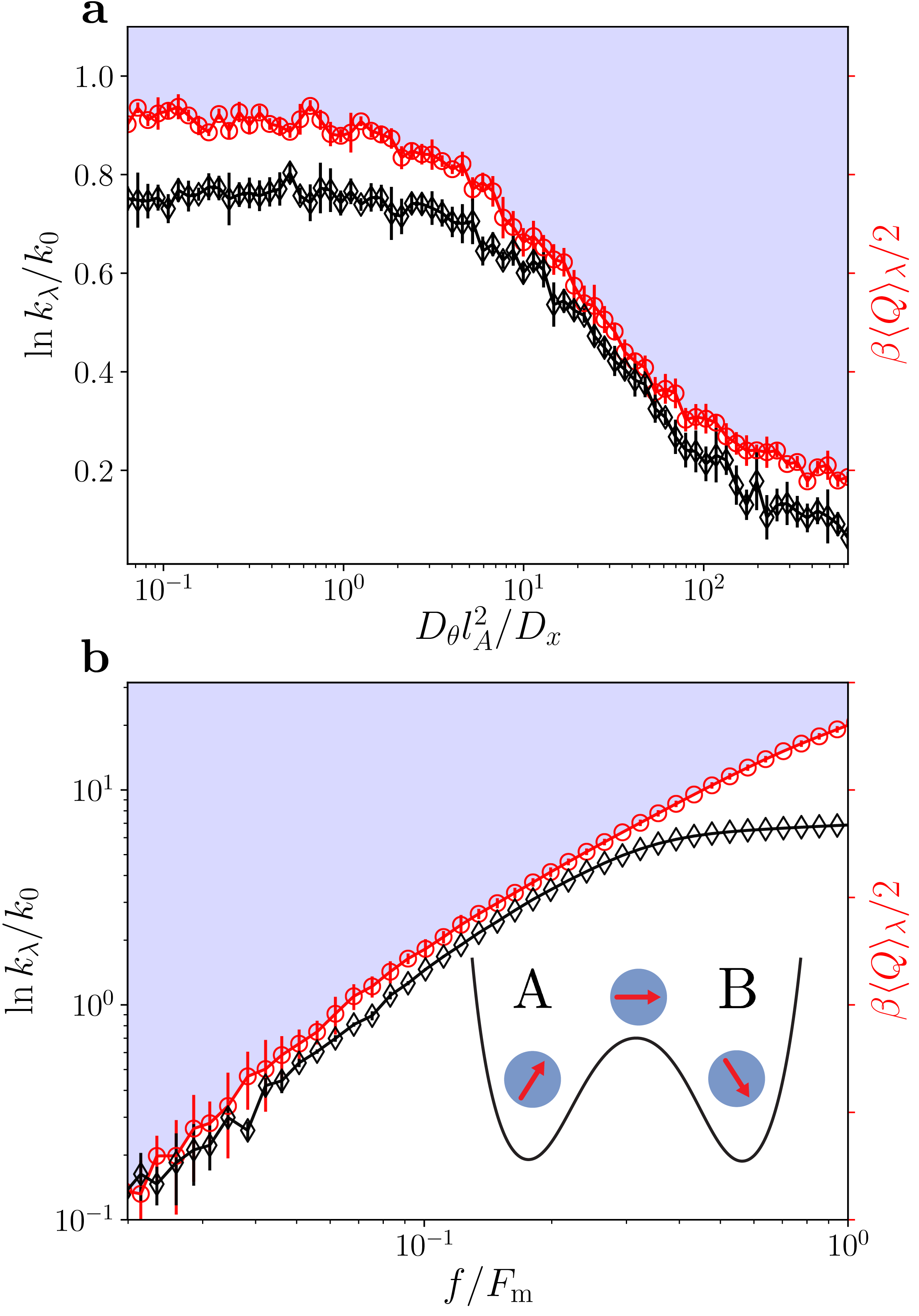} 
\caption{\label{fig:3} Rate enhancement for an active Brownian particle. \textbf{a}, Rate enhancement as a function of rotational diffusivity constant $D_\theta l_A^2/D_x$ for $f = 1$.  \textbf{b}, Rate enhancement as a function of the magnitude of active driving relative to the maximum force opposing the transition in equilibrium $f/F_m$ for $D_\theta=1/2$. In both, the rate enhancement (black diamonds) is bounded by the dissipated heat (red circles). }
\end{figure}

Figure~\ref{fig:3}a) shows the dependence of the rate enhancement on the rotational diffusivity, $D_\theta$ at fixed $f=1$. The small $D_\theta$ limit is a quasi-stationary regime corresponding to an equilibrium system with an additional linear force added to the potential. 
%
%
Increasing the rotational diffusivity decreases the persistence of the driving, and as a consequence $\langle Q \rangle_\lambda$ and $\kampo$ fall off in this limit. In the large $D_\theta$ limit, the system is effectively in equilibrium at an elevated temperature, as $\lambda$ averages to 0. Lower rate enhancement and little dissipation are observed across this range of $D_\theta$, and our bound is uniformly close. These results are consistent with a recent study in which an effective potential approach was used to derive $\kampo$ for an active Ornstein–Uhlenbeck process in a cubic well\cite{active_escape_PRE}.  As shown in Fig.~\ref{fig:3}b), our bound is closest to the true rate enhancement when driving is small compared to the maximum force needed to surmount the barrier in equilibrium,  $f < F_\mathrm{m}$ where $F_m=8 \Delta V_A/3\sqrt{3 l_A}$. In that regime, the heat and rate enhancement both scale with $f^2$, as predicted by linear response theory. When the protocol and gradient forces are comparable, the transition ceases being a rare event and further increasing $f$ has little effect on the rate, but increases the heat.

As a second test case, we consider underdamped dynamics with time-periodic driving, $p=1$. This model, known as the Duffing oscillator,\cite{SR_Hanggi}
is the simplest model of a stochastic pump and one whose nonequilibrium behavior is marked by significant nonlinearity. As an underdamped process, its barrier crossing behavior is determined both by spatial as well as energy diffusion, in which both position and velocity correlations play a role. The equation of motion is given by $m \ddot x   = -\gamma \dot x - \partial_x V(x) + \lambda(t) + \sqrt{2 \kB T} \eta_x$, where the mass $m$ reflects the change to underdamped dynamics.  Again we take a symmetric potential, $l_{A}=l_{B}=1$, now with $\beta \Delta V_A=\beta \Delta V_B=14$ and $m=\beta = \gamma= D_x = 1$. 

\begin{figure}[t]
\includegraphics[width=.925\linewidth, height=1.36\linewidth]{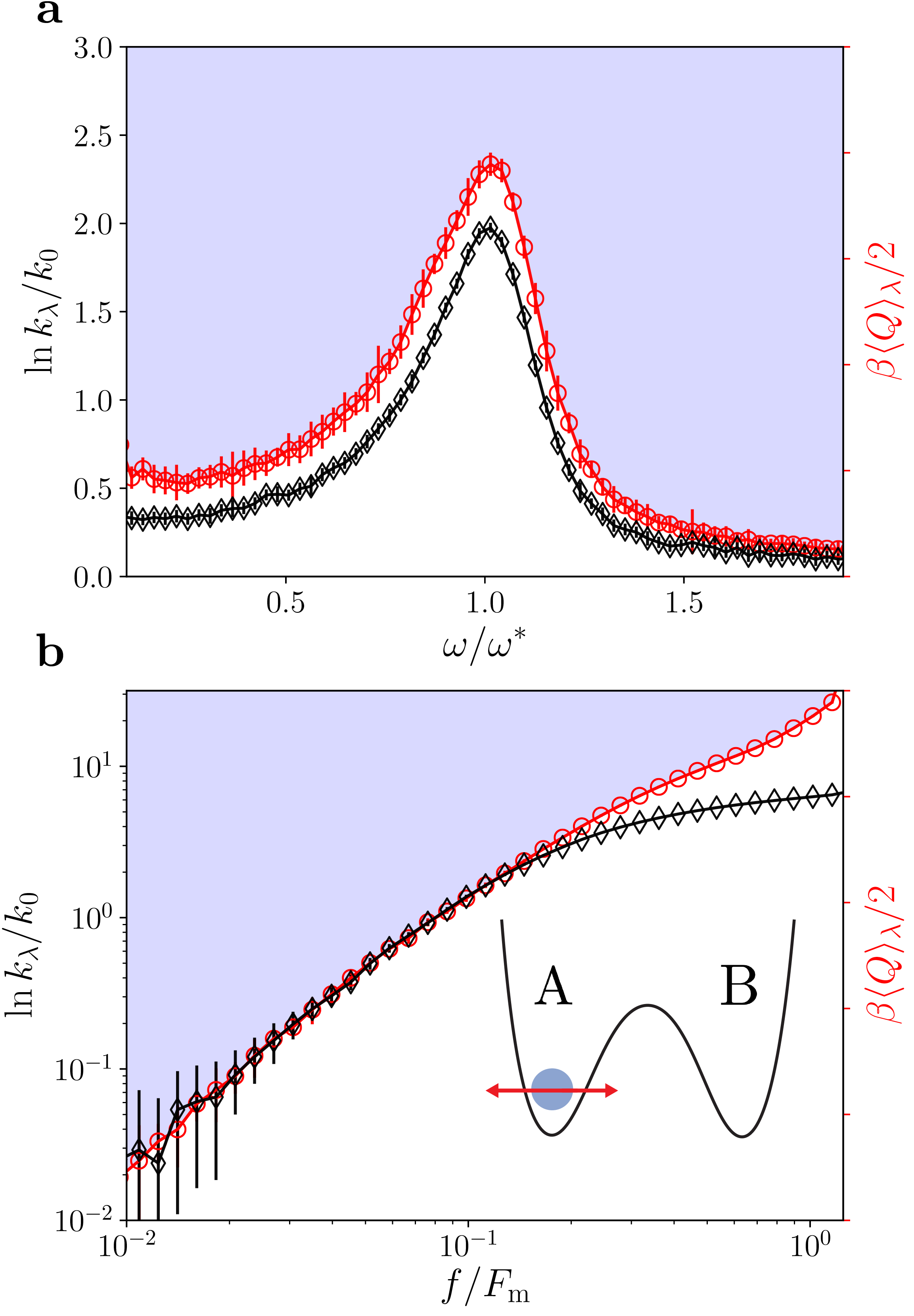} 
\caption{\label{fig:2} Rate enhancement in the Duffing oscillator.  \textbf{a}, Rate enhancement as a function of the magnitude of driving frequency relative to the natural frequency of the equilibrium system, $\omega/\omega^*$, for $f=1.4$. \textbf{b}, Rate enhancement as a function of the magnitude of function of the periodic driving relative to the maximum force opposing the transition in equilibrium $f/F_m$ for $\omega/\omega^*=1$ at $f=1.4$.  In both, the rate enhancement (black diamonds) is bounded by the dissipated heat (red circles).
}
\end{figure}

Figure~\ref{fig:2}a) shows that for a moderate force, $f/F_\mathrm{m} \approx 0.13$, there is an optimal driving frequency, denoted here as $\omega^{*}$, which greatly enhances the transition rate. This phenomenon is known as stochastic resonance\cite{SR_Hanggi}. 
For slow driving, $\omega \ll \omega^{*}$, the particle typically makes a transition before the external force reaches its maximum. For  $\omega \gg \omega^{*}$ driving is inefficient, and on average requires multiple forcing cycles before presenting a chance to cross the barrier with the help of a positive force within the time of a typical transition. The approximate shape of the rate enhancement profile is Lorentzian, a trait inherited from the absorption lineshape of an underdamped harmonic oscillator. In this case, the resonant frequency coincides with the curvature of the equilibrium double-well potential driven with a quasi-static force, $\omega^{*} \approx \sqrt{8(\Delta V-f)-\gamma^2/4}$\cite{jung1993}. We find near saturation of the bound throughout a wide range of frequencies and across even such nonlinear behavior as stochastic resonance. In Fig.~\ref{fig:2}b), we plot $\kneq/\keq$ and $\beta \langle Q \rangle_\lambda/2$ against the driving amplitude relative to $F_\mathrm{m}$. As in the other examples, the bound on rate enhancement is tight so long as the metastability of state $A$ is preserved. 

Thus far, we have characterized rates with transition paths, which lend themselves to a natural response theory for $\ln k_\lambda$, and therefore the ratio of rates. However, the survival probability $p_\lambda(t) =\exp(- k_\lambda t)  =1- Z_{AB}(\lambda)/ Z_{A}(\lambda)$ contains similar information in the case of two metastable states, when $h_A+h_B = 1$. Falasco and Esposito recently\cite{Diss_time_TUR} worked with this quantity to prove a speed limit on escape processes.  Extending these results, we derive analogous bounds on absolute reaction rates (Sec. S4). Assuming rare rates, we bound the ratio of survival probabilities $p_\lambda(t)/p_0(t)$, and thus the difference of rates $\kneq-\keq$, from above and below using the relative action and Jensen's inequality. The final result reads  
\begin{equation} \label{survive}
\beta \left \langle  \Delta \dot{U}_\lambda \right \rangle_\lambda \leq  k_\lambda-k_0 \leq  \beta \left \langle  \Delta \dot{U}_\lambda \right \rangle_0
\end{equation}
where the rate of change of the relative action forms an envelope around the change in the transition rate. If $k_0 \ll k_\lambda$ then Eq.~\ref{survive} acts as a speed limit on the forced process. On the other hand, if $k_\lambda \ll k_0$, Eq.~\ref{survive} reports on the minimum dissipation required to slow down a fast equilibrium process. 
%
The envelope in Eq. \ref{survive} is similar to the bound derived by Falasco and Esposito in that it relates differences of rates to conditioned path ensemble averages. However, in Eq. \ref{survive}, two forward rates under different dynamics are considered, while in Ref.~\cite{Diss_time_TUR} a forward rate is compared to its backward counterpart in the same ensemble. By considering a pair of rates explicitly related by time reversal symmetry, the result of Falasco and Esposito follows directly from the fluctuation theorem\cite{CrooksFT}. By contrast, the results presented in this work do not follow from the fluctuation theorem; rather, they are a result of trajectory ensemble reweighting.

The variational relationship in Eq.~\ref{diss_bound} between the transition rate and a thermodynamic quantity is reminiscent of equilibrium transition state theory, where the rate is bounded by the thermal flux times the probability that a thermal fluctuation brings the system from an initial reactant state to a rare transition state. From the Kawasaki distribution,\cite{CrooksFT} the nonequilibrium configurational
distribution is related to the equilibrium one by a cumulant generating function of the excess dissipation conditioned on ending at a specific configuration. As explained in the Sec. S5, neglecting changes to the flux generated by coupling to a nonequilibrium process, the transition state theory estimate of the rate is bounded by the average heat conditioned on ending at the transition state. Eq. ~\ref{diss_bound} thus has the interpretation of a nonequilibrium extension to standard transition state theory, and is expected to be a good approximation to the rate enhancement in cases where the system spends little time at the top of the barrier. 

While our examples have focused on simple one dimensional models, our formalism is general and can be straightforwardly applied to many-body interacting systems. One immediate consequence of higher dimensionality is that unlike in the examples explored here, applied forces need not be aligned with the direction of the most likely transition path. In such cases, we expect the efficiency with which an arbitrarily applied force enhances the rate to be suppressed relative to the largely saturated bounds we have found here, as energy may be transduced into modes not correlated with overcoming the barrier. In light of our results, a natural optimal control problem arises in which nonequilibrium protocols can be constructed that minimize the dissipation for a given desired rate enhancement. Methods to perform such optimizations employing molecular simulation and importance sampling have been recently developed and show promise in complex systems \cite{Touchette_control,Avishek, Todd_near_optimal_control}.
Similarly, the protocols uncovered by such an optimization have the potential to lend mechanistic insight into reactions far from equilibrium, as basic concepts like free energy barriers and gradient flows cease being well defined. The effect of dissipation on rate enhancement under counter-diabatic~\cite{cd_evo} constraints, as well as in discrete-state networks and reaction-diffusion settings, remains to be seen.   

Overall, our investigations build a general framework for the systematic and computationally efficient characterization of rate enhancement. Predicting how structure and external influence conspire to alter reaction rates far from equilibrium is of immediate importance in designing proteins, enzymes, small-molecule drugs, and the complex environments in which they operate. We foresee future studies in interacting colloidal and polymeric systems, both in shear and confining geometries that change dynamically in time \cite{netz_expir}. Applications to heterogeneous growth, nucleation and jamming will also be interesting avenues to explore \cite{oskar_jamming}. Time-dependent chemical potential gradients in gated release and receptor binding contexts, as well as designing interaction protocols \cite{DNA_interactions} for quick and robust self-assembly \cite{self_assembly_microgears} and pattern-formation \cite{Denk_Frey} are another set of pressing examples to which our theory can apply. Local heating in ATP hydrolysis and facilitated diffusion on DNA\cite{MMR_slideing}, where electric fields play a pivotal role, are all more complex problems that may prove fruitful to study in this manner.

\section*{Methods} 
In our first main text example, we consider a double-well potential $V(x)$ pieced together continuously at $x = 0$, and the parameters dictating its shape and the form of the driving are drawn uniformly. This is done $n$ times to create a parameter database (each entry labeled by integer $i$ from $0$ to $n-1$) before any simulations are run. Each point in Fig. 1, corresponds to a unique set of system parameters chosen as follows. Fixing $\beta  = D_x = 1$, we uniformly draw $\beta \Delta V_A, \beta \Delta V_\mathrm{B} \in (3,7)$, $ l^2_\mathrm{i} \in 8 \Delta V_\mathrm{i}(\omega^{-2}_\mathrm{max},\omega^{-2}_\mathrm{min})$, $p\in [0,1]$, $\log f \in [-1,\log f_\mathrm{ max}]$, and $D_\theta \in  [0.1,10]$, where we constrain the natural frequencies of the wells to lie between $\omega_\mathrm{min} = 3.5$ and $\omega_\mathrm{min} = 7.5$. 

Simulations for the overdamped systems studied, including the mixed driving system and the active barrier crossing, are propagated with a first order stochastic Euler integrator using a timestep of $\Delta t=10^{-2}$ in units of the $A$-state relaxation time $\tau_A = l_A/\sqrt{8 \Delta V_A}$. For the underdamped Duffing oscillator calculations, we used a OVRVO\cite{OVRVO} integrator, employing a symplectic 5-step Strang splitting of the time-evolution operator with the same time-step. Before turning $\lam$ on at time zero, the dynamics are first evolved in equilibrium, $\lam(t) = 0$, for $t = 1.5 \times 10^3 \Delta t$, to equilibrate. We calculate $k_\lam$ and $k_0$ by counting the number of transitions from $A$ to $B$ and dividing by simulation time, $1.25-5 \times 10^7 \Delta t$. Transitions are counted when $x(t)$ passes from $A$ to $B$ and stays there at least order $2.75 \Delta t$ steps without recrossing, which we find to yield consistent results with rates extracted from the side-side correlation function. We perform $24$ simulations, each of which yields a noisy estimate of the rate, and we record the mean and standard error bars.

We compute the time-dependent average $\left \langle  Q \right \rangle_\lambda(t)$, independently of the rates, as follows. Using the same criteria for a transition as defined in the proceeding paragraph, we perform as many simulations as is needed to ensure sufficient statistics, which, depending on the specific system parameters, corresponds to $10^{3-4}$ transitions. At each point in time, we record the instantaneous rate of heat dissipation as well as an indicator function to coarse-grain $x(t)$ into either state $A$, state $B$, or the transition state, which in this example we define as $x\in A \cup B$, the intersection between states. Immediately after each simulation, we collect and save a list of times when transitions started, $t_1$, and ended, $t_2$. Once all simulations are complete, we histogram transition path times, $t_2-t_1$.  We integrate over a discretized version of the heat flux in Eq. \ref{Q_heat}, $Q(t = n \Delta t) = \int_0^t dt' \lam(t') \dot x(t') \approx \sum^{n-1}_{i = 0} \lam_i ( x_{i+1}-x_i)$.
In order to account for all transitions, we choose an initial observation time $t$ equal to that required for $99\%$ of transition paths to proceed start to finish. For each transition, we position a window of length $t$ so that it ends immediately after $t_2$, and integrate over it. We slide this window to the right by $O(10 \Delta t)$ steps and integrate again, repeating until the endpoints of the window . The coarse-grained time $ \approx \Delta t$ controls how highly correlated reactive trajectories are. Choosing it is thus some what of an art in the sense that when very small, one must integrate over many more paths, but if it is too large, $\left \langle  Q \right \rangle_\lambda$ will have greater statistical error. If, at any point, a window passes through a transition region without seeing a transition, due to coarse-graining, the observation time is increased by a small factor of $t \mapsto t (1+1/4)$ and the integration is restarted from the beginning. 

The rates are small quantities, prone to high statistical uncertainty which is amplified when the log of their ratio is taken. Because of this, we rerun all points that have greater standard error, in either enhancement or dissipation, than the mean. The majority or these points correspond to \textit{lightly} forced systems where the maximum protocol amplitude is usually much smaller than the thermal energy scale, $\beta |\lam| \sim 10^{-1}$. In this regime, $\ln k_\lam/k_0 \ll 1$, is well within linear response. Since we are focused on the case where the equilibrium rate changes appreciably, and know that by time-reversal symmetry arguments, the only contribution in this near-equilibrium case comes from dissipation, throughout the main text we only show points with $\ln k_\lam/k_0$ larger than a small number, which we choose to be $10^{-2}$.

{\bf Acknowledgments:} This material is based upon work supported by the National Science Foundation under grant no. CHE-1954580 and by a Scialog Program sponsored jointly by Research Corporation for Science Advancement and the Gordon and Betty Moore Foundation.


\section*{References} 

%

\end{document}


\beginsupplement
\title{Supplemental Material: \\
Dissipation bounds the amplification of transition rates far from equilibrium}

\preprint{APS/123-QED}
\author{Benjamin Kuznets-Speck}
\affiliation{Biophysics Graduate Group, University of California, Berkeley, CA, 94720, USA \looseness=-1}
\author{David T. Limmer\email{dlimmer@berkeley.edu}}%
\affiliation{Chemistry Department, University of California, Berkeley, CA, 94720, USA \looseness=-1}
\affiliation{Chemical Sciences Division, Lawrence Berkeley National Laboratory, Berkeley, CA, 94720, USA \looseness=-1}
\affiliation{Material Sciences Division, Lawrence Berkeley National Laboratory, Berkeley, CA, 94720, USA \looseness=-1}
\affiliation{Kavli Energy NanoSciences Institute, University of California, Berkeley, CA, 94720, USA \looseness=-1}

\date{\today}


\maketitle


\thispagestyle{empty}
\onecolumngrid


\section{Generalized bound on rate enhancement}
Our main results were obtained under the assumption that the probability of starting in $A$ does not change much under the influence of driving $Z_A(\lam) \approx Z_A(0)$. In this section, we derive a more general bound that relaxes this assumption, and discuss in what cases we expect initial conditions to play a significant role. The factor $Z_A(0)/Z_A(\lam)$ in question is the ratio of single-time probabilities and has played an important role in the development of Monte-Carlo sampling on the space of driving protocols.\cite{Todd_design_space} As with the transition path partition function, it is still possible to express such a ratio in terms of the moment generating function of the relative action 
\begin{equation}
    \frac{Z_A(0)}{Z_A(\lam)} = \left \langle \frac{\rho_0}{\rho_\lambda} e^{-\beta \Delta U_\lambda} \right \rangle_{A,\lambda} = \left \langle \frac{\rho_\lambda}{\rho_0} e^{\beta \Delta U_\lambda} \right \rangle_{A,0}^{-1},
\end{equation}
which implies 
\begin{equation}
      -\beta \langle \Delta U_\lambda \rangle_{A, \lam}-D_\mathrm{KL}(\rho_\lambda||\rho_0) \leq \ln \frac{Z_A(0)}{Z_A(\lam)} \leq   -\beta \langle \Delta U_\lambda \rangle_{A, 0}+D_\mathrm{KL}(\rho_\lambda||\rho_0) 
\end{equation}
where the paths of any length are conditioned to start in state $A$. After restoring initial conditions, Eq. 9 of the main text then implies  
\begin{equation}
      \beta (\langle \Delta U_\lambda \rangle_0- \langle \Delta U_\lambda \rangle_{A, \lam}) -  J(\rho_\lam||\rho_0) \leq \ln \frac{k_\lam}{k_0} \leq   \beta (\langle \Delta U_\lambda \rangle_\lam-\langle \Delta U_\lambda \rangle_{A, 0}) + J(\rho_\lam||\rho_0), \label{gen_bound}
\end{equation}
which is our most general result. Initial conditions  $\rho_\lam(x_0)$ and $\rho_0(x_0)$ enter in through a symmetric variant of the Kullback–Leibler divergence known as the Jeffreys divergence, 
\begin{equation}
    J(\rho_\lam||\rho_0) =  D_{KL}(\rho_\lam||\rho_0) + D_{KL}(\rho_0||\rho_\lam) =   \int dx_0 (\rho_\lam - \rho_0)\ln \frac{\rho_\lam}{\rho_0} \geq 0,
\end{equation}
and we will discuss shortly under what conditions it can be neglected. Even when its most likely state changes, if $A$ remains metastable, the system will start in an effective stationary state with flux and force that both vanish on average, leaving the negative, $\lambda^2$ piece of the dynamical activity as the only portion of $\langle \Delta U_\lambda \rangle_{A, 0}$ left. However, this piece does not depend on the trajectory since under our assumptions, the external force is independent of the system configuration, so it cancels in the relative action with an identical contribution from the driven ensemble. We can therefore deduce that only the trajectory dependent portions of  $\langle \Delta U_\lambda \rangle_\lambda$, and $J(\rho_\lambda||\rho_\lambda)$ are likely to contribute to our upper bound. 

To understand the influence of initial conditions, we begin by noting that they cease to enter into the bound when the initial configuration is deterministic or the system is prepared according to the same distribution, either $\rho_\lambda$ or $\rho_0$. Next, that $\rho_\lambda$ and $\rho_0$ are equivalent in linear response means $J(\rho_\lambda||\rho_\lambda)$ vanishes in this regime. Thus, if making the transition is correlated with the initial distribution in state $A$, then whenever that correlation remains relatively invariant under driving, initial conditions can be neglected. Perhaps most relevant to the present work and the study of rare events  is the situation when initial conditions becoming uncorrelated with the 'bulk' of the transition path, and are forgotten. The reason why loosing memory of initial conditions is far from exotic for such reactive trajectories stems directly from the fact that state $A$ is assumed, by definition of the rate constant, to be metastable, harboring a local minima of the free energy landscape. The initial portion of a typical rare event transition path from $A$ to $B$ will therefore relax exponentially quickly in this local basin and loose memory over a short timescale $\tau_A$ determined by the curvature of the basin. Since the rate constant is defined for times that are order $\tau_A$, as detailed in the paragraph following Eq. 5 in the main text, the brunt of reactive paths are marked by boundary portions of rapid relaxation on either side of the transition region. When state $A$ is suitably defined so that this memory loss occurs in both the driven and reference ensembles, all that matters is that the particle starts somewhere in state $A$, which is guaranteed by the transition portion of the path probability. In other words, knowledge about the character of state $A$ is contained directly in $\langle \Delta U_\lambda \rangle_\lambda$ when both rate constants are defined with a common observation time, which is the topic of Sec. S3. When the driving is so extreme that a common observation time ensuring relaxation before escaping state $A$ cannot be obtained, the rate constant effectively looses meaning. Such atypical cases can be addressed taking into account initial conformational distributions, or preferably, redefining $A$ so that $k_0$ and $k_\lambda$ are comparable by memory-less transition paths of the same length.               

\section{Conditions for neglecting the dynamical activity}
In order to understand when the contribution to the dynamical activity to the rate enhancement bound can be neglected, we consider both simplified limiting cases that are analytically tractable, as well as additional numerical experiments on the systems considered in the main text. In general, the dissipative bound is valid in cases where the activity can be neglected due to its size or if it can be shown to be strictly negative. The former case can be argued to occur near-equilibrium or when specific spatial symmetries exist that result in its average being small. The latter case occurs when a particle traverses  both large, narrow barriers as well as broad diffusive barriers. 

\subsection{Limiting cases of high symmetric and broad diffusive barriers}

For simplicity and concreteness, we consider an overdamped particle in a $1d$ symmetric double well potential. The equation of motion is analogous to that considered in the first two examples in the main text,
\begin{equation}
    \gamma \dot x (t) = F_\lambda(x)+ \sqrt{2 \kB T\gamma} \eta, \quad \quad  \langle \eta(t) \rangle =0 \quad \quad  \langle \eta(t)\eta(t') \rangle = \delta(t-t'), 
\end{equation}
where $F_\lambda(x) = - V'(x) + \lambda(t)$ is the total force, including the gradient and non-gradient contributions.
From the equation of motion, it is straight-forward to show that the conditioned transition probability takes the form  $\ln P_\lambda[x(t)|  x(0)]  = \beta U_\lam[x(t)] + C$, where $C$ is a constant and the Onsager-Machlup path-action is
\begin{equation}
   U_\lam = -\frac{1}{4 \gamma} \tint dt' \, \left \{ \gamma \dot{x}(t') - F_\lambda[x(t')] \right \}^2
\end{equation}
for an ensemble with finite nonequilibrium driving and path of length $t$. 

In the limit that the barrier separating two metastable states is large, such that transitions between them are rare, the rate can be computed by extremizing the path action, $\delta U_\lam/\delta x =0$. Preforming the functional differentiation, the instantonic trajectory satisfies the second order differential equation
\begin{equation}
\gamma^2 \ddot{x} -  \dot{F}_\lambda[x(t')] - F_\lambda[x(t)] \partial_x F_\lambda[x(t)]  =0,
\end{equation}
which can, in principle, be solved subject to the boundary conditions of starting in state $A$ and ending in state $B$.  Here we will consider both states being defined at specific points $x=x_A$ and $x=x_B$ for states $A$ and $B$, respectfully. However, solving this equation is difficult for nonconservative potentials and time dependent driving forces. We consider the case of a large, sharply peaked barrier, such that the maximum force due to the potential, $F_m$, is much larger than the magnitude of the applied driving force, $f=\max |\lambda(t)| \ll F_m$. In this limit, the instanton equation simplifies to
\begin{equation}
\gamma^2 \ddot{x} \approx   V'(x)V''(x) 
\end{equation}
which results in the trajectory traced out by the particle in equilibrium. Its first integral is a constant of motion, yielding
\begin{equation}
\gamma^2 \dot{x}^2 =   [V'(x)]^2,
\end{equation}
that provides both branches of the instanton, or extremal path.\cite{Boucher2} The positive root yields the trajectory beginning at $x_A$ and ending at the maximum of the potential separating the two metastable states, $x=x_m$, and results in a positive contribution to the action. The negative root yields the trajectory beginning at $x_m$ and ending at $x_B$, with zero action. The resultant total action is approximated in this limit as
\begin{equation}
   U_\lam \approx  \int_0^{t/2} dt' \,   \dot{x}(t') F_\lambda[x(t')]=  \int_0^{t/2} dt' \,   \dot{x}(t') \lam(t') - \int_{x_A}^{x^\ddagger} dx \,  V'(x)
\end{equation}
where consistent with the assumption of an equilibrium trajectory minimizing action, we have neglected terms of order $O(\lambda^2)$ that are strictly negative and invoked time reversal symmetry to set the domain of the integrals. The first term is the equilibrium change in energy, while the second is the heat. Using this result, we can compute the averaged relative action between equilibrium and nonequilibrium path ensembles
\begin{equation}
   \langle \Delta U_\lam \rangle_\lambda \approx \QnAB/2
\end{equation}
which has no contributions from the activity and coincides with our main result. This calculation clarifies a relevant linear response limit in which the dissipative bound is tight. It is one in which the typical reactive trajectory in and out of equilibrium are similar, following a gradient path, which occurs in cases where the conservative forces experienced during a transition are large relative to the nonequilibrium driving forces, $f\ll F_m$. 

For a second example, we consider a barrier that is locally parabolic in the vicinity of its maximum,
$$
V(x) \approx -k x^2/2 + V_0
$$
where $k$ denotes its local curvature and $V_0$ is the offset from the minimum in the $A$ state. In the limit that $\beta V_0 \gg1$, we can assume that the majority of the action required to overcome the barrier is localized to the region around the maximum. In such a case, the stochastic action in the presence of the external force is extremized by the solution of 
\begin{equation}
\gamma^2 \ddot{x}(t)   =k^2 x(t) + k \lambda(t) +  \gamma \dot{\lambda}(t)   \, ,
\end{equation}
for the instantonic trajectory subject to boundary conditions which we take to be symmetric about the maximum, $x(0)=-x_0$ and $x(t^*)=x_0$. It is convienent to introduce the charactoristic relaxation time, $\tau= \gamma/k$. This linear ordinary differential equation can be solved by the method of Laplace transforms, yielding
\begin{equation}
x(t) =  -x_0 \cosh(t/\tau)+\frac{x_0 +x_0 \cosh(t^*/\tau)}{ \sinh(t^*/\tau)} \sinh(t/\tau) + f(t)/\gamma -\frac{f(t^*)/\gamma}{ \sinh(t^*/\tau)} \sinh(t/\tau)
\end{equation}
where $f(t)$ is the convolution of the external force with the Green's function,
\begin{equation}
f(t) = \int_0^t dt' \lambda (t') e^{(t-t')/\tau}
\end{equation}
and acts as an inhomogenious source. From the definitions in the main text, the heat and activity are given by
\begin{equation}
Q= \int_0^{t^*} dt  \, \dot{x}(t) \lambda(t)   \quad \quad \Gamma= -\frac{1}{2\gamma} \int_0^{t^*} dt \,  \lambda(t) \left [ 2 k x(t)  +\lambda(t) \right ]
\end{equation}
whose averages are computed within the instantonic trajectory. The heat has a familiar form. The activity includes a contributions due to the product of the gradient force and the external force, and a contribution due to the external force squared. While the latter is strictly negative, path independent, and can be dropped while still satisfying the bound, the former may be positive or negative. 

In the limit of small $\lambda(t)$, it can be verified that the first contribution to the activity vanishes, as the instanton trajectory spends equal time on the left and right side of the barrier with corresponding equal and opposite forces from the external potential. The second term enters proportional to $-\lambda^2(t)$ which can also be neglected if $\lambda(t)$ is small. This is identical to the near equilibrium case considered above. Analogously, in the limit that $\dot{\lambda}\approx 0$, such that over the transition the external force is well approximated by a constant, by symmetry the activity will be strictly negative. Consider for concreteness a periodic force $\lambda(t)= f \cos(\omega t)$ with characteristic amplitude $f$ and frequency $\omega$. In the limit that $\omega \tau \ll 1$, the heat will be  $\QnAB=2 f x_0$ and the activity $\AnAB=-f^2 t^*/2 \gamma<0$. In the opposite limit that $\omega \tau \gg 1$, many cycles of the driving force will elapse during the instantonic trajectory and its influence on breaking the symmetry around the barrier will vanish. In the case of a monochromatic driving force, the heat is  $\QnAB=f^2 t^*/2 \gamma$ and the activity $\AnAB=-f^2 t^*/4 \gamma< 0$ can be neglected in the bound. Away from these cases, the trajectory need not be symmetrically distributed around the barrier, and the activity maybe finite. However, this occurs when the driving is large, in which case the second term in the activity will dominate leading to it being negative.

Finally, in the case where the barrier is broad and flat, we can approximate the motion at the top as free diffusion conditioned a set of starting and ending points, $x_A$ and $x_B$. Specifically, when $V(x)=0$,  for an arbitrary external force, the average heat and activity simplify to
$$
\QnAB= \frac{\left ( x_B - x_A \right )}{t^*} \int_0^{t^*} dt  \, \lam(t)+
\gamma^{-1} \int_0^{t^*} dt  \, \left [ \lam(t)-\bar{\lam}(t^*) \right]^2
 \quad \AnAB= -\frac{1}{2\gamma} \int_0^{t^*} dt \,  \lambda^2(t)
$$
where $\bar{\lam}(t^*)$ denotes the external force averaged over the transition time $t^*$. The activity is manifestly negative and can be dropped from the dissipative bound. As above, it is negligible when the scale of $\lambda$ is small as it scales quadratically. Taking $|x_B - x_A|$ and $t^*$ large, the heat becomes the displacement times the average force, $\QnAB \approx \left ( x_B - x_A \right ) \bar{\lam}$ and the activity $\AnAB \approx - t^* \bar{\lambda^2} /2\gamma$ where $\bar{\lambda^2}$ is the average squared size of the external driving.

\subsection{Relative magnitudes of the heat and activity in the transition path ensemble}

\begin{figure}
 \includegraphics[width=.85\linewidth]{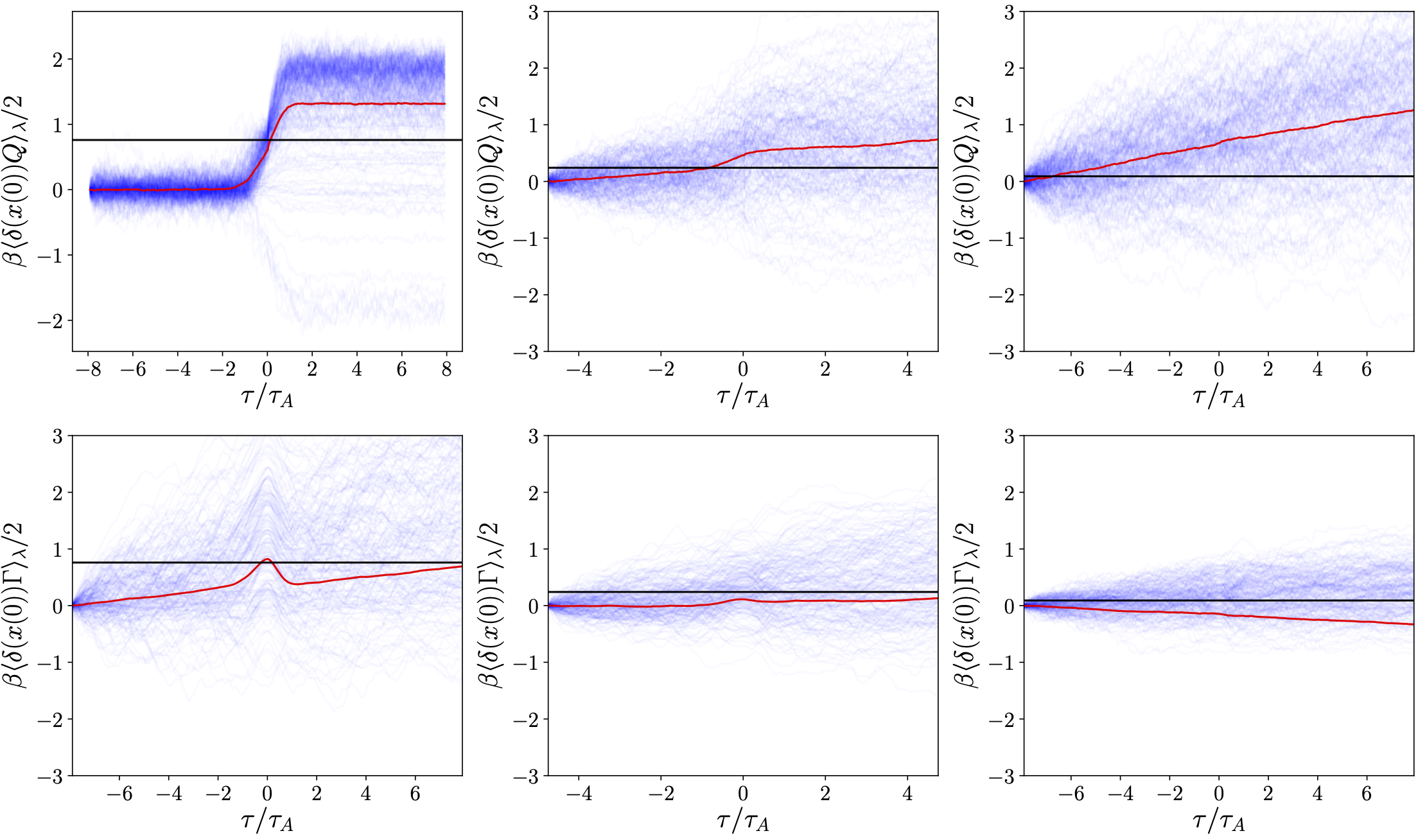}
 \caption{\label{fig:2} 
 Additional examples of active barrier crossing. Rate amplification $\ln k_\lam/k_0$ (black, from counting transitions),  dissipation (red, $\mathbf{top~row}$), and dynamical activity (red, $\mathbf{bottom~row}$) along the typical reaction trajectory, in units of the reactant (state $A$) relaxation time $\tau_A$. All parameters except for active diffusivity $D_\theta$ relative to that at which the enhancement is half its maximum $D_\theta^*$ are set according to Fig. 3 a). $\mathbf{Left}$: slow driving with a long persistence time $ D_\theta/D^*_\theta \approx 2 \times 10^{-3}  $.  $\mathbf{Center}$: close half-maximum persistence $D_\theta/D^*_\theta \approx 1 $. $\mathbf{Right}$: low-persistence forcing $D_\theta/D^*_\theta \approx 7$.}
 \vspace{1cm}
 \includegraphics[width=.85\linewidth]{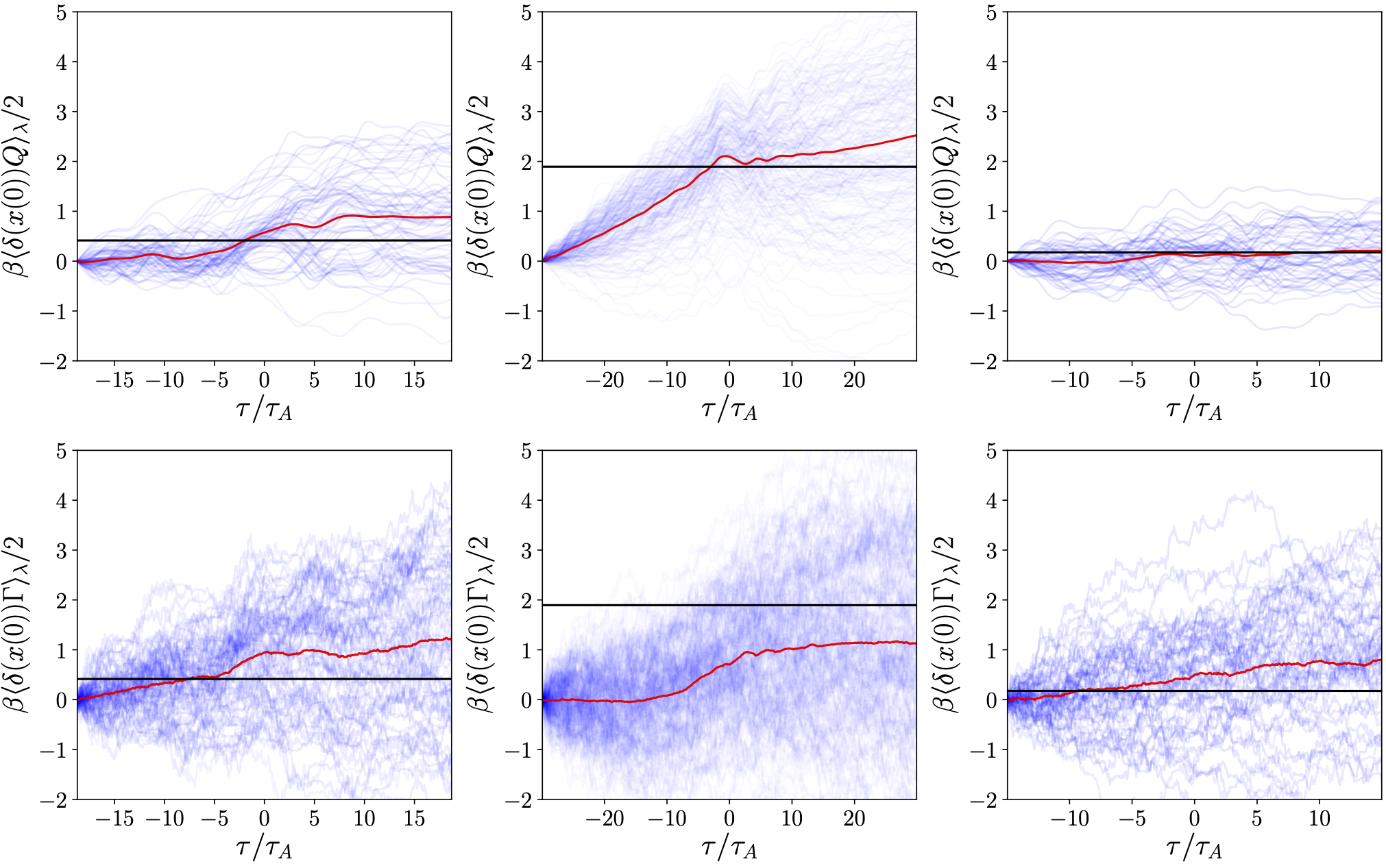}
 \caption{\label{fig:3} 
 Revisiting stochastic resonance. Rate amplification $\ln k_\lam/k_0$ (black, from counting transitions),  dissipation (red, $\mathbf{top~row}$), and dynamical activity (red, $\mathbf{bottom~row}$) along the typical reaction trajectory, in units of the reactant (state $A$) relaxation time $\tau_A$. Parameters apart from driving frequency $\omega$ are set according to FIG. 3 a) in the main text. $\mathbf{Left}$: slow, quasi-adiabatic driving $\omega/\omega^* \approx 0.42 $.  $\mathbf{Center}$: close to resonance $\omega/\omega^* \approx 1.04 $. $\mathbf{Right}$: very fast forcing $\omega/\omega^* \approx 1.55 $.}
 \end{figure}

Here, we discuss the results of additional simulations on systems featured in the main text Figs. 2 and 3. Shown in  Fig.~\ref{fig:2} and \ref{fig:3} are the heat (top row), and also the activity (bottom row) \emph{throughout} the course of the transition. Each column represents a driving \emph{speed}: low, medium and, high, from left to right, and we fix all other parameters as in the main text. The active Brownian particle in Fig.~\ref{fig:2} is driven at a speed $D_\theta$ equal to its inverse auto-correlation time, and $\omega$ is the analogous variable in the Duffing Oscillator depicted in The quantities plotted in these figures  as a function of time, in units of the reactant ($A$) relaxation time, are averages of $\int^{\tau-t/2}_{-t/2} \dot Q$ and $\int^{\tau-t/2}_{-t/2} \dot \Gamma$, which are \emph{not} part of the transition path ensemble until crossing occurs at time zero. For the overdamped active Brownian particle, $\dot \Gamma = -\gamma^{-1} \lambda(t) F_\lambda[x(t)]$ and in the underdamped Duffing oscillator $\dot \Gamma = \gamma^{-1} \lambda(t)(\gamma \ddot x(t) - F_\lambda[x(t)])$, which can be derived by constructing the symmetric part of the corresponding path action. 

As the reaction proceeds from beginning to end, heat and activity are computed in the path ensemble defined by the additional constraint that the particle position coincides with the transition state, $x^\ddag$ at time zero. That is, we average over all trajectories starting at time $-t/2$ in $A$ before crossing at time $0$ and ending up in $B$ a time $t/2$ after that. 
Since both the underlying system and the driving are symmetric in time in the long-time limit, it is reasonable to believe the instanton connecting $A$ to $B$ is also, so this scheme should sample the $AB$ ensemble as the number of trajectories becomes large. 
We collect long trajectories of $\dot Q$ and $\dot \Gamma $, integrating from $-t/2$ to $\tau-t/2$ for $\tau \in (0,t)$. Specifically, we run simulations of $\sim 10^{8-9} \Delta t$ until $\sim 10^{3-4}$ reactions are observed. 
The majority of paths lie above $\QnAB$ but there are a number of outlying negative heat trajectories, wherein the particle crosses the barrier while also \emph{opposes} the driving force, which is a consequence of the integral fluctuation theorem. 

In Heat (top, red) rises past the rate enhancement (black) in the immediate vicinity of this time, an observation consistent with the non-equilibrium transition-state theory we develop below. On the other hand, the activity need not serve as an upper bound, as is the case for particles driven at the half-maximum speed and resonance frequency in in the center column of Fig.~\ref{fig:2} and \ref{fig:3}, respectively. Like we discussed in the section prior, the activity accumulated over a symmetric barrier should vanish by symmetry in the adiabatic limit, a signature which is approximately realized in Fig.~\ref{fig:2} (bottom left). Finally, when driving varies so quickly that it couples effectively as a second bath, Fig.~\ref{fig:2}, the activity is negative, as would be expected for free diffusion.

\section{Separation of timescales in the driven and equilibrium ensembles}
The existence of a time-independent transition rate between two metastable states requires a separation of timescales between local relaxation within a metastable state and the characteristic time to transition between the two states. \cite{TPSrate2} The rate enhancement relation in the main text additionally requires that separation exists for both the driven and equilibrium ensembles, and further that these two intervals defined by the local relaxation and typical transition waiting times have some amount of overlap, starting at $t_\mathrm{min}$. 

Crucially, the observation time $t$ past which our bounds are defined and valid in the conditioned path ensemble must be chosen at least as large as the minimum overlap time $t > t_\mathrm{min}$. Before $t_\mathrm{min}$, heat will increase approximately linearly. An observed transition to different behavior, typically occurring around the reactant relaxation time $\tau_A$, can be employed as a signature of when our bound is tightest in situations wherein the true rate enhancement is unknown. 

\begin{figure}[ht]
\includegraphics[width=.9\linewidth]{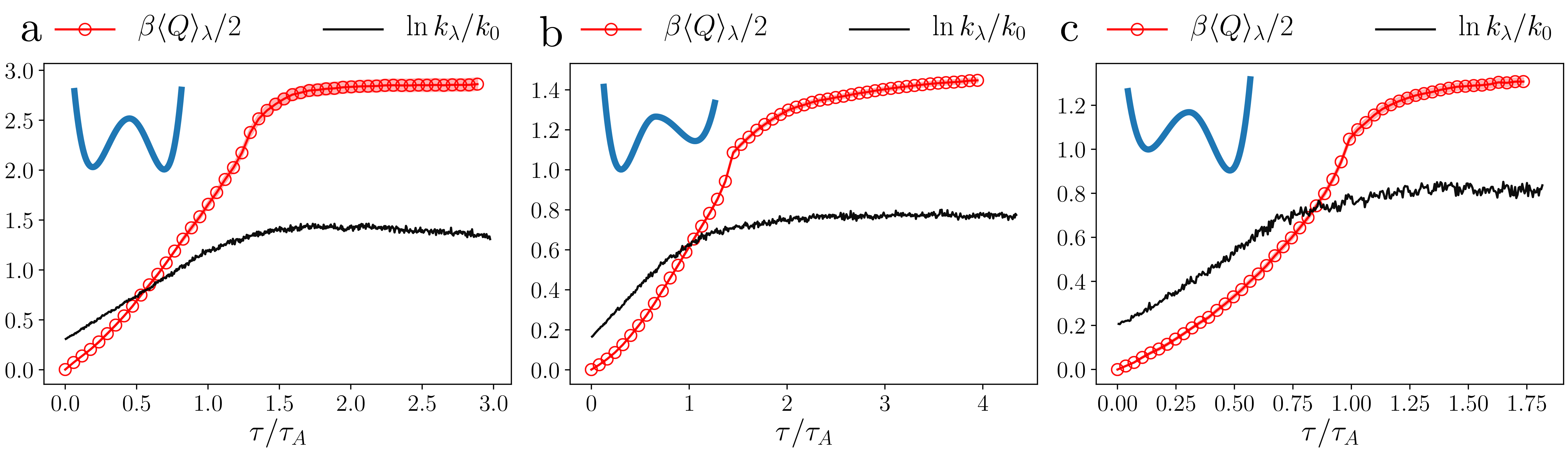}
\caption{\label{fig:1} 
Rate amplification $k_\lam/k_0$ (black) and dissipation (red) for three examples of heterogeneously driven two-state systems. The shape of the equilibrium potential for each system is inset (blue) in the upper-left hand corner, and time is taken in units of the relaxation time for state $A$ (the state on the left side of each inset). $\mathbf{a}$, transitions in an approximately symmetric double-well driven by equal parts deterministic and active forces, amplitudes summing to around half the equilibrium well-depth. $\mathbf{b}$ Escape from a basin of attraction to a less-stable state driven by mostly active forces with total max-amplitude around $0.35 \Delta V_A$. $\mathbf{c}$ Excursions to a state with greater stability, forces mainly by a time-periodic protocol with maximum amplitude about $0.75 \Delta V_A$.}
\end{figure}

Heat does not accumulate in a quasi-equilibrium state, so if both $A$ and $B$ remain deeply metastable, one can expect there to exist an interval of time starting around $t_\mathrm{min}$ when the particle commits to state $B$ and $\QnAB$ varies relatively slowly. To test these predictions, we consider the mixed driving system studied in Fig. 1. Specifically, Fig.~\ref{fig:1} shows the log ratio of the rates in and out of equilibrium as a function of time for that system, together with the accumulated heat, where $\tau$ is an intermediate time between 0 and the observation time $t$. In all cases where there are is large barrier separating states $A$ and $B$, and the applied force $f$ is smaller than the maximum force $F_m$ due to the potential, the ratio of rates plateaus within a time of $\mathcal{O}(1)$. Further, over the time window when the rate ratio is time independent, heat gains support at a much smaller rate than prior to this window, plateauing in the case of dual metastability in Fig.~\ref{fig:1} a. Therefore, under these conditions, we find path ensemble averages in question to be somewhat insensitive to precise time at which they are taken. 

\section{Bounds on survival probabilities}
The recently proposed dissipation-time uncertainty relation \cite{Diss_time_TUR2} claimed that the rate of steady state entropy production bounds the difference of forward to backward, denoted here with a tilde, transition rates, out of and into a metastable state $A$, from above
\begin{equation}
    k_\lambda-\tilde{k}_\lambda\leq  \beta  \langle \dot{Q}   \rangle_\lam.
\end{equation}
In this section, we follow the same line of thinking, but use the definition of $\Delta U$ in place of the traditional fluctuation theorem.
For a \emph{two-state system}, the indicator functions (Main: Eq. 2) defining states $A$ and $B$ are simply related by $h_A(t)+h_B(t) = 1$. In this case, the cumulative probability $p(t)$ that no transitions occur up to time $t$, the survival probability, is related to the probability that at least one occurs in the same way: 
\begin{equation}
   p_\lam(t) = \frac{Z_{AA}(\lam, t)}{Z_A} = 1-\frac{Z_{AB}(\lam, t)}{Z_A}. 
\end{equation}
From Eq. 1. in the main text, this relation implies that the transition rate is given by $k_\lam = -d \ln p_\lam(t)/dt$. Bounding $p_\lam(t)/p_0(t)$ from above and below in the $AA$ path ensemble, and subsequently using the fact that $h_A(t) = 1-h_B(t)$ to convert to the $AB$ ensemble yields the envelope presented in Eq. 13. For example,  
\begin{equation}
   \frac{p_0}{p_\lam} = \frac{Z_{AA}(0, t)}{Z_{AA}(\lam, t)} = \langle e^{-\beta \Delta U} \rangle_{AA,\lambda} \geq  e^{-\beta \langle \Delta U \rangle_{AA,\lambda}}  = e^{-\beta (1-\langle \Delta U \rangle_{AB,\lambda})} \implies k_\lam - k_0 \geq \beta \langle \Delta \dot{U}_\lam \rangle_{AB, \lambda}
\end{equation}
yields an upper bound, where, to be explicit, we label averages by which path ensemble they belong to ($AA$ and $AB$).
Again, bounds in Eq. 13 assume the meta-stability of the $A$ state is sufficiently preserved, though this can be relaxed by following the procedure laid out in SM section 3. We leave it to future work to delve into the utility of bounding the change of the rate $k_\lam - k_0$ in this fashion.

\section{Connection to transition-state theory}
Our main focus is on rates calculated within the transition path ensemble, obtained by approximating the sum over time-dependent trajectories. The benefit of this approach is that all that one needs is a definition of two states, $A$ and $B$. Detailed mechanistic knowledge of the kinetic bottlenecks through which the transitions pass is unnecessary, alleviating the potentially hard problem of finding a good reaction coordinate, $q$, and relevant transition state, $q^\ddagger$, dividing metastable states $A$ and $B$. However, if a pertinent reaction coordinate is known, the rate can be estimated by much simpler means using transition-state theory, which trades in the path partition function $Z_{AB}(\lam)$ for the average flux through a transition state times the probability of being at $q^\ddagger$. \cite{TST2} Generalizing the traditional result, for a nonequilibrium system,
\begin{equation}
    k_\lam \leq \frac{\langle |\dot{q} |\rangle_{\lam,q^\ddagger} }{2 } \frac{\rho_\lam(q^\ddagger)}{\rho_\lam(A)}.
\end{equation}
where $\langle |\dot{q} |\rangle_{\lam,q^\ddagger}$ is the average flux through the transition state, and $\rho_\lam(q^\ddagger)/\rho_\lam(A)$ is the relative probability of finding the system at the transition state relative to the total probability of finding the system in state $A$. In equilibrium, the equipartition theorem dictates that the average flux is independent of $q$, but that is not necessarily true away from equilibrium. The transition-state theory approximation is an underestimate of the true rate, as it neglects those trajectories that switch direction after already passing $q^\ddagger$, and can be made exact by computing a correction termed the transmission coefficient. 

For rare, instantonic transitions, the relative probability to reach the transition state, $\rho_\lam(q^\ddagger)/\rho_\lam(A)$, is by far the dominant contribution to the transition-state theory rate. In equilibrium, this term gives rise to the Arrhenius law 
\begin{equation}
   k_0 \propto \frac{\rho_0(q^\ddagger)}{\rho_0(q_A)}  = e^{-\beta \Delta F}
\end{equation}
 where $\Delta F = F(q^\ddagger)-F(q_A)$ is the height of the free energy barrier separating $q^\ddagger$ from the most probable state $q_A  \in A$. Out of equilibrium, rate estimation via transition state theory becomes a much more difficult problem, because while a time averaged stationary distribution $\rho_\lam(q)$ still exists, one cannot in general express it in a simple close form. 
 
However, using the Kawasaki distribution equation, we can approximate the nonequilibrium steady-state distribution in terms of a product of the equilibrium distribution and a correction dependent on the mean dissipation. As formulated by Crooks \cite{CrooksFT} for stochastic dynamics and similar to that derived by Evans and Searles\cite{Evans_Searles}, the Kawasaki distribution provides a relation for the probability of being in state $q$ having started in an equilibrium distribution, and cumulant generating function of the accumulated dissipated heat. Specifically,
\begin{equation}
\rho_0(q) =\rho_\lam(q)\left  \langle e^{-\beta Q} \right \rangle_{\lam,q}
\end{equation}
where $Q$ is the heat dissipated to the environment, $\rho_0(q)$ is the initial equilibrium probability of state $q$, $\rho_\lam(q)$ is the nonequilibrium probability of state $q$, and the brackets $\langle \dots  \rangle_{\lam,q}$ denote an average under the driving force $\lam$ conditioned on ending at state $q$. Taking the saddle point, and applying Jensen's inequality,
\begin{equation}
\label{Eq:Kaw}
\frac{\rho_\lam(q)}{\rho_0(q)} \le e^{\beta \left  \langle  Q \right \rangle_{\lam,q}}
\end{equation}
we arrive at a bound for the ratio of the nonequilibrium to equilibrium distributions.

The bound of the steady state distribution cannot be applied directly to the estimation of rates, but under mild assumptions it can provide an estimate similar to the result in the main text. To proceed, we first assume that the probability of being in state $A$ is unchanged between the equilibrium and nonequilibrium ensemble, $\rho_\lam(A) = \rho_0(A)$. From the Jarzynski equality \cite{Jarzynski_equality} and integral fluctuation theorems\cite{Gallavotti_Cohen}, this is true if the protocol is cyclic, because then $\langle  \exp(\beta Q)  \rangle_\lam=1$. This can be obtained by taking the observation time  long enough for the driven system to relax back to the original equilibrium distribution by the final time, since in that case, the conservative heat and the change in free energy both cancel, leaving the excess heat behind. Assuming like initial distributions is approximately valid in the limit that $A$ is deeply metastable. Next, we assume that we can ignore the change to the flux over the transition state due to coupling to $\lam$. Moreover, the fluctuation-response inequality\cite{Sasa_FRI} implies the first order change to the flux is expected to scale as $\sim \sqrt{  \beta \langle  Q  \rangle_{\lam}}$, which is subdominant to the exponential dependence from the change in the probability distribution. Finally, we assume that the transmission coefficient in and out of equilibrium are the same. This is likely a good assumption in the limit that the original transition state theory estimate in equilibrium is tight, and the transition is instantonic such that the protocol is slowly varying relative to the typical transition path time. Under those assumptions, we find
\begin{equation}
\frac{k_\lam}{k_0} \lesssim e^{\beta \left  \langle  Q \right \rangle_{\lam,q^\ddagger}}
\end{equation}
where we have applied Eq.~\ref{Eq:Kaw} to the probability of reaching the transition state. Note here, as previously, the activity does not enter the approximate bound, and we have a purely mechanical relationship between the enhancement speed of a process and the energy required.  Further, the dissipated heat is that accumulated in going from state $A$ to the transition state, $q^\ddagger$. In an instantonic limit, and near equilibrium, we expect the heat accumulated in reaching the top of the barrier to be half that to reach state $B$, which would result in an analogous expression as the main text.





%